\documentclass[twocolumn,showpacs,superscriptaddress,aps,prl]{revtex4-1}
\usepackage{bm}
\usepackage{amssymb,amsmath}
\usepackage{graphicx}
\usepackage{color}

\begin{document}
\title{Quantum critical behavior of a three-dimensional \\ superfluid-Mott glass transition}
\author{Jack Crewse}
\author{Cameron Lerch}
\author{Thomas Vojta}
\affiliation{Department of Physics, Missouri University of Science \& Technology, \\ Rolla, MO, 65409, USA}

\begin{abstract}
The superfluid to insulator quantum phase transition of a three-dimensional particle-hole symmetric system of disordered bosons is studied. To this end, a site-diluted quantum rotor Hamiltonian is mapped onto a classical (3+1)-dimensional XY model with columnar disorder and analyzed by means of large-scale Monte Carlo simulations. The superfluid-Mott insulator transition of the clean, undiluted system is in the 4D XY universality class and shows mean-field critical behavior with logarithmic corrections. The clean correlation length exponent $\nu = 1/2$ violates the Harris criterion, indicating that disorder must be a relevant perturbation. For nonzero dilutions below the lattice percolation threshold of $p_c = 0.688392$, our simulations yield conventional power-law critical behavior with dilution-independent critical exponents $z=1.67(6)$, $\nu = 0.90(5)$, $\beta/\nu = 1.09(3)$, and $\gamma/\nu = 2.50(3)$. The critical behavior of the transition across the lattice percolation threshold is controlled by the classical percolation exponents. Our results are discussed in the context of a classification of disordered quantum phase transitions, as well as experiments in superfluids, superconductors and magnetic systems. 
\end{abstract}

\maketitle

\section{Introduction}
\label{sec:Intro}
Models of disordered and interacting bosons can be employed to describe a wide variety of
physical phenomena including  helium absorbed in porous media \cite{CHSTR83,Reppy84},
superconducting thin films \cite{HavilandLiuGoldman89,HebardPaalanen90}, Josephson
junction arrays \cite{ZFEM92,ZEGM96}, ultracold atoms in disordered optical
lattices \cite{WPMZCD09,KSMBE13,DTGRMGIM14}, and certain disordered quantum magnets
\cite{OosawaTanaka02,HZMR10,Yuetal12,Huevonenetal12,ZheludevRoscilde13}.

It is well established \cite{GiamarchiSchulz88,FisherFisher88,FWGF89} that the Mott-insulating
and superfluid phases of these models are always separated by an insulating ``glass'' phase in
which rare large regions of local superfluid order (superfluid ``puddles'') coexist with the insulating bulk.
The glass phase thus acts as a Griffiths phase \cite{Griffiths69,ThillHuse95,Vojta06,Vojta10}
of the superfluid-Mott insulator quantum phase transition.

The nature of the glassy intermediate phase depends on the qualitative properties of the disorder.
For generic disorder (realized, e.g., via a random potential for the bosons), it is the
so-called Bose glass, a compressible gapless insulator. The zero-temperature phase
transition between the superfluid and Bose glass ground states has recently reattracted
lots of attention as new analytical \cite{WeichmanMukhopadhyay07}, numerical
\cite{PCLB06,MeierWallin12,NgSorensen15,ALLL15,YCKP14}, and experimental \cite{Yuetal12,Yuetal12prb,Huevonenetal12}
work has challenged the scaling relation \cite{FisherFisher88,FWGF89}  $z=d$
between the dynamical exponent $z$ and the space dimensionality $d$ as well as
the value of the crossover exponent $\phi$ that governs the shape of the finite-temperature phase boundary.

If the system is particle-hole symmetric even in the presence of disorder, the intermediate
phase between superfluid and Mott insulator is not a Bose glass but
the \emph{incompressible} gapless Mott glass \cite{GiamarchiLeDoussalOrignac01,WeichmanMukhopadhyay08}.
(This state is sometimes called random-rod glass because in a classical representation the disorder
takes the form of infinitely long parallel rods.)
The zero-temperature phase transition between the superfluid and Mott glass ground states
has received less attention than the Bose glass transition, perhaps because in some experimental
applications the condition of exact particle-hole symmetry is hard to realize and
requires fine tuning. Note, however,
that the particle-hole symmetry appears naturally in magnetic realizations of disordered boson physics
due to the up-down symmetry of the spin Hamiltonian in the absence of an external magnetic field.

We have recently determined the quantum critical behavior of the superfluid-Mott glass transition
in two space dimensions using large-scale Monte-Carlo simulations \cite{Vojtaetal16}, resolving earlier
contradicting predictions in the literature \cite{ProkofevSvistunov04,IyerPekkerRefael12,SLRT14}.
However, magnetic realizations of the Mott glass state have mostly been observed in
\emph{three-dimensional} disordered magnets. To the best of our knowledge, quantitative results
for the three-dimensional superfluid-Mott glass transition do not yet exist.

To investigate this transition, we analyze a site-diluted three-dimensional quantum
rotor model with particle-hole symmetry. We map this quantum Hamiltonian onto
a classical $(3+1)$-dimensional XY model with columnar defects. We then carry out
Monte Carlo simulations for systems of up to 56 million lattice sites, averaging each data set
over 2500 to 20,000 disorder configurations. For dilutions $p$ below the lattice percolation threshold
$p_c\approx 0.688392$,\cite{BFMPR99} we find the superfluid-Mott glass quantum phase transition to be characterized
by universal (dilution-independent) critical exponents. The dynamical exponent takes the value $z=1.67(6)$,
and the correlation length exponent is $\nu=0.90(5)$, fulfilling the inequality $\nu> 2/d$.\cite{Harris74,CCFS86}
For the order parameter exponent $\beta$ and the susceptibility exponent $\gamma$, we find
$\beta/\nu=1.09(3)$ and $\gamma/\nu=2.50(3)$, respectively. This gives an anomalous dimension of $\eta=-0.50(3)$.
These exponents fulfill the hyperscaling relation $2\beta/\nu+\gamma/\nu=d+z$.
As a byproduct, our simulations also yield the critical behavior of the clean (undiluted) four-dimensional
XY model with high accuracy. It is characterized by mean-field exponents with logarithmic corrections
(as expected at the upper critical dimension) and agrees well with the predictions of a generalized scaling theory\cite{Kenna04}.

Our paper is organized as follows. Section \ref{sec:Theory} defines
the three-dimensional quantum rotor Hamiltonian and the quantum-to-classical mapping to a $(3+1)$-dimensional
classical XY model. It also introduces our finite-size scaling technique (that does not require prior knowledge
of the dynamical exponent) as well as the generalized scaling theory \cite{Kenna04} for the clean case.
Monte Carlo results for the clean and disordered phase 
transitions are presented in Sec. \ref{sec:MC}.
We summarize and conclude in Sec. \ref{sec:Conclusions}.

\section{Theory} 
\label{sec:Theory}
\subsection{Diluted Rotor Model}
\label{sec:Model}
We investigate the superfluid-Mott glass transition by means of a site-diluted quantum rotor model residing on a three-dimensional cubic lattice,  
\begin{equation}
\label{Hq}
H = \frac{U}{2}\sum_i\epsilon_i(\hat{n}_i-\bar{n}_i)^2 - J\sum_{\langle ij \rangle}\epsilon_i\epsilon_j \cos(\hat{\phi}_i - \hat{\phi}_j),
\end{equation} 
where $\hat{n}_i$, $\bar{n}_i$, $\hat{\phi}_i$ are the number operator, offset charge, and phase operator of site $i$, respectively. $U$ and $J$ represent, respectively, the charging energy and Josephson junction coupling of the sites. We define the dilution, or impurity concentration, as the probability $p$, that a site is vacant. The independent quenched, random variables $\epsilon_i$ then take on the values 0 (vacancy) with probability $p$ and 1 (occupied site) with probability $1-p$. 

The superfluid and Mott glass states can be modeled by this Hamiltonian when considering a particle-hole symmetric system with offset charges $\bar{n}_i$ = 0 and commensurate (integer) fillings $\langle \hat{n}_i \rangle$. The phase diagram of this Hamiltonian has been extensively studied \cite{FWGF89,WeichmanMukhopadhyay07}. For dominant charging energy, $U \gg J$, the ground state is a Mott insulator. For dominant Josephson junction coupling, $J \gg U$, the ground state of the system instead becomes a superfluid. Of course, this behavior is only relevant for dilutions below the lattice percolation threshold, $p_c \approx 0.688392$. Dilutions above $p_c$ cause the lattice to break down into disconnected finite-size clusters, preventing the establishment of any long-range ordered superfluid phase. Between the superfluid and Mott insulator phases a third, intermediate phase emerges. In our particle-hole symmetric case, this is the Mott glass, an incompressible, gapless insulator.  The quantum phase transition from the superfluid to the Mott glass state is the focus of the present investigation. A detailed discussion of these phases and their properties can be found, e.g., in Ref. \onlinecite{WeichmanMukhopadhyay08}.
 
\subsection{Quantum-to-Classical Mapping}
\label{sec:QCmap}
\begin{figure}
\includegraphics[width=0.75\columnwidth]{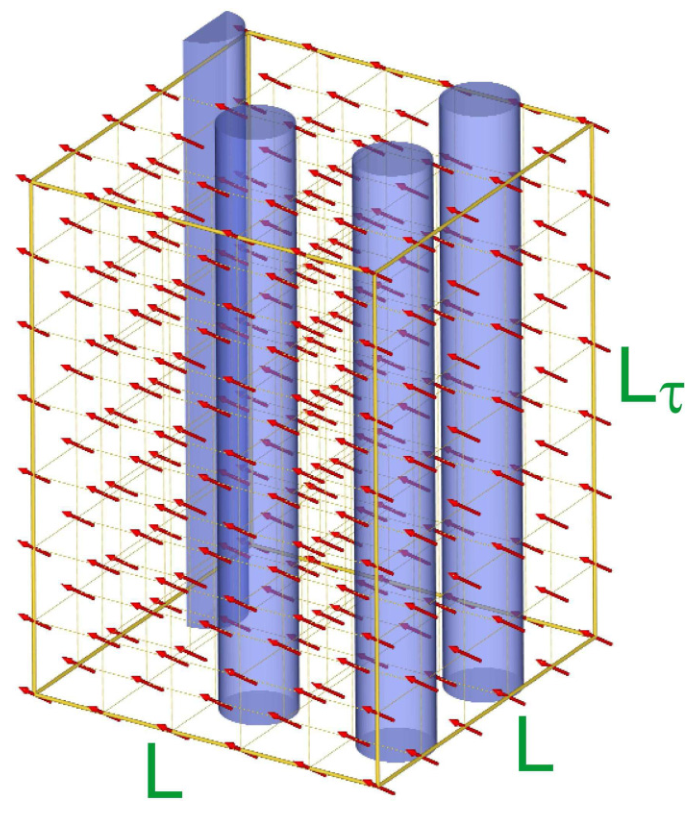}
\caption{(2+1)-dimensional analog of the system (\ref{Hc}). Arrows are the classical spins $\mathbf{S}$. Columns represent the site-vacancies perfectly correlated in imaginary-time. A true sketch of the system (\ref{Hc}) would be four-dimensional with vacant 'columns' in the imaginary-time dimension.}
\label{fig:swisscheese}
\end{figure}

As we are interested only in universal properties of the transition, we simplify our study of the critical behavior by mapping the 3-dimensional quantum Hamiltionian (\ref{Hq}) onto a classical Hamiltonian of total dimensionality $D = d+1 = 4$.\cite{WSGY94} The mapping gives (see Fig. \ref{fig:swisscheese})
\begin{equation}
\label{Hc}
H_{cl} = - J_s \sum_{\langle ij \rangle,\tau} \epsilon_i\epsilon_j \mathbf{S}_{i,\tau}\cdot \mathbf{S}_{j,\tau} - J_{\tau} \sum_{i,\tau}\epsilon_i\mathbf{S}_{i,\tau}\cdot \mathbf{S}_{i,\tau+1}
\end{equation}
with $\mathbf{S}_{i,\tau}$ being an O(2) unit vector at space coordinate $i$ and imaginary-time coordinate $\tau$. Within this mapping, the "classical" temperature $T$ of the Hamiltonian (\ref{Hc}) does not refer to the physical temperature of the quantum system (which is zero at the quantum phase transition). Instead, the constants $J_s/T$ and $J_{\tau}/T$ that appear in the classical partition function represent the coupling constants $J$ and $U$ of the quantum system, and the "classical" temperature is used to tune the couplings and drive the system through the transition. Additionally, the expected universality of the critical behavior allows us to ignore the exact numerical values of $J_s$ and $J_{\tau}$, so we set $J_s = J_{\tau} = 1$ in the following.

\subsection{Clean (undiluted) Critical Behavior}
\label{sec:CleanTheory}
In the clean limit $p=0$ (no vacancies) the Hamiltonian (\ref{Hc}) becomes isotropic in the space and imaginary time dimensions, thus simplifying the system to the four-dimensional classical XY model. This places the clean system at the upper-critical dimension $D_c^+=4$ of the XY universality class. Renormalization group calculations have shown that the transition at $D_c^+$ exhibits  mean-field critical behavior with logarithmic corrections to scaling\cite{Kenna04}. These calculations yield a scaling form for the free energy 
\begin{equation}
f_L(r,H) = L^{-4}\mathcal{F}\Big(rL^2(\ln L)^{1/10},HL^3(\ln L)^{1/4}\Big)
\label{eqn:clean_free}
\end{equation}
where $r = (T-T_c)/T_c$ and $H$ represent the reduced temperature and field conjugate to the order parameter, respectively. Appropriate derivatives of $f_L(r,H)$ yield the dependencies of the order parameter $m$ and its susceptibility $\chi$ on the system size $L$ at criticality
\begin{align}
m \propto L^{-1}(\ln L)^{1/4} \label{cleanmscaling}\\
\chi \propto L^2(\ln L)^{1/2}. \label{cleanchiscaling}
\end{align}
This implies $\beta/\nu = 1$ and $\gamma/\nu = 2$ for the order parameter and susceptibility critical exponents, respectively. The correlation length exponent can also be extracted via the quantity $d(\ln m)/dT$, which from (\ref{eqn:clean_free}) leads to the scaling form
\begin{equation}
\frac{d(\ln m)}{dT} \propto L^2(\ln L)^{1/10} \label{cleandlnmdtscaling}
\end{equation}
implying a correlation length exponent $\nu = 1/2$. This value, however, violates the Harris criterion\cite{Harris74} for stability of phase transitions against weak disorder, $d\nu > 2$, where $d=3$ is the number of dimensions with randomness, i.e., the space dimensionality. Thus the clean XY critical point is unstable against the columnar defects we introduce. As a result, we expect the diluted system to exhibit new critical behavior and exponents. 

\subsection{Anisotropic finite-size scaling}
\label{sec:AFSS}
Variables of scale dimension zero are especially useful in the determination of a system's critical behavior within the framework of finite-size scaling\cite{Barber_review83}.  For example, central to our study is the Binder cumulant
\begin{equation}
\label{Binder}
g_{av} = \left[1-\frac{\langle |\mathbf{m}|^4 \rangle}{3\langle |\mathbf{m}|^2 \rangle^2}\right]_{\rm dis}
\end{equation}
where $\mathbf{m} = (1/N)\sum_{i,\tau}\mathbf{S}_{i,\tau}$ is the order parameter ($N$ being the total number of lattice sites of the classical Hamiltonian (\ref{Hc})). Additionally, $\langle ... \rangle$ denotes the Monte Carlo average, and $[...]_{\rm dis}$ an average over disorder configurations. In the thermodynamic limit, $g_{av}$ is expected to take the value $2/3$ in the superfluid phase and the value $1/3$ in both the Mott glass and Mott insulator phases. We also study the correlation lengths in the space and imaginary-time directions \cite{CooperFreedmanPreston82,Kim93,CGGP01} 
\begin{align}
\label{spaceCorrLength}
\xi_s    = & \Bigg[\bigg(\frac{\tilde{G}(0,0) - \tilde{G}(q_{s0},0)}{q_{s0}^2\tilde{G}(q_{s0},0)}\bigg)^{1/2}\Bigg]_{\text{dis}}, \\
\label{timeCorrLength}
\xi_{\tau} = & \Bigg[\bigg(\frac{\tilde{G}(0,0) - \tilde{G}(0,q_{{\tau}0})}{q_{{\tau}0}^2\tilde{G}(0,q_{{\tau}0})}\bigg)^{1/2}\Bigg]_{\text{dis}} 
\end{align}
where $\tilde G(q_{s0},q_{\tau 0})$ is the Fourier transform of the spin-spin correlation function, $q_{s0}$ and $q_{\tau 0}$ are the minimum wavelengths in the space and imaginary-time directions, respectively. 

For an isotropic system of system size $L$, and distance $r = (T-T_c)/T_c$ from criticality, the Binder cumulant has the finite-size scaling form $g_{av}(r,L) = X(rL^{1/\nu})$. This guarantees that at $r=0$, the $g_{av} \text{ vs } r$ plots for different system sizes will cross at a value $g_{av}(0,L) = X(0)$, allowing us to easily locate $T_c$. However, the introduction of quenched disorder in the space dimensions breaks the isotropy between space and imaginary time, thus requiring us to distinguish the system sizes $L$ in the space direction and $L_{\tau}$, in the imaginary-time direction.

The finite-size scaling form of the Binder cumulant now depends on the relation between $L$ and $L_{\tau}$. For conventional power-law scaling it reads
\begin{equation}
\label{BinderPowScaling}
g_{av}(r,L,L_{\tau}) = X_{g_{av}}(rL^{1/\nu},L_{\tau}/L^z)
\end{equation}
where $z$ is the dynamical exponent, whereas for activated scaling the term $L_{\tau}/L^z$ in (\ref{BinderPowScaling}) is replaced by $\ln(L_{\tau})/L^{\psi}$ with $\psi$ the tunneling exponent. A classification scheme based on the dimensionality of locally ordered rare regions in the disordered system suggest that we should expect power-law scaling \cite{Vojta06,VojtaHoyos14}. Rare region dimensionality for our XY model is $d_{RR}=1$ (infinitely extended rare regions in the single imaginary-time direction). The lower critical dimension of the XY model is $D_c^-=2$, thus we have $d_{RR}<D_c^-$. This puts the system (\ref{Hc}) firmly into class A of the classification implying power-law dynamical scaling\cite{VojtaHoyos14}. This also means that our system is not expected to display power-law Griffiths singularities. Instead observables such as the order parameter susceptibility $\chi$ show conventional behavior. Specifically, $\chi$ will remain finite in the Mott glass phase, and rare regions make exponentially small contributions.

For anisotropic systems, we must modify our approach to finite-size scaling. Due to our initial ignorance of the dynamical exponent $z$, we do not know the appropriate sample sizes $L \times L_{\tau}$ to fix the second argument of the scaling function (\ref{BinderPowScaling}) in the simulations. We can take advantage of some of the Binder cumulant's properties to find the appropriate ratios (``optimal shapes") of $L_{\tau}/L$ and thus our dynamical exponent $z$.\cite{GuoBhattHuse94,RiegerYoung94,SknepnekVojtaVojta04} For a fixed spatial size $L$, $g_{av}$ as a function of $L_{\tau}$ will exhibit a maximum at the point $(L_{\tau}^{\text{max}},g_{av}^{\text{max}})$. At this point the ratio $L_{\tau}/L$ behaves like the corresponding ratio of correlation lengths $\xi_{\tau}/\xi_s$ and designates the ``optimal shape" for that given $L$. For values of $L_{\tau}$ above or below the maximum, the system can be decomposed into independent blocks which decreases the value of $g_{av}$. At criticality $L_{\tau}^{\text{max}}$ is proportional to $L^z$. Samples of optimal shape thus fix the second argument of the scaling form (\ref{BinderPowScaling}), allowing one to carry out the rest of the finite-size scaling analysis as usual.  

Actually carrying out the calculations requires an iterative approach. An educated guess is made for an initial value of the dynamical exponent $z$ (e.g. the value calculated for the (2+1)d case)\cite{Vojtaetal16}. The (approximate) crossings of the $g_{av}$ vs $r$ curves  for samples of the resulting shapes give an estimate for $T_c$. The temperature is then fixed at this estimate of $T_c$ and $g_{av}$ as a function of $L_{\tau}$ is analyzed. The points of maximum value $g_{av}^{\text{max}}$ at $L_{\tau}^{\text{max}}$ can then be calculated and give improved estimates for the optimal shapes and thus an improved estimate on $z$. For $T>T_c$ the $g_{av}^{\text{max}}$ values will tend towards their disordered (decreasing) values with increasing system size, for values $T<T_c$ they tend towards their ordered (increasing) values for increasing system size. Thus, for a given estimate for $T_c$, the trends of $g_{av}^{\text{max}}$ with system size allow us to determine how to adjust our $T_c$ estimate for the next iteration. Using this procedure the values of $T_c$ and $z$ converge quickly, requiring only about 3-5 iterations. 

Once we have determined the value of $z$ for the system, the usual finite-size analysis can be carried out with the scaling forms 
\begin{equation}
m = L^{-\beta/\nu}X_m(rL^{1/\nu},L_{\tau}/L^z) \label{mScaling}
\end{equation}
\begin{equation}
\chi = L^{\gamma/\nu}X_{\chi}(rL^{1/\nu},L_{\tau}/L^z) \label{chiScaling}
\end{equation}
where $\beta$ and $\gamma$ are the order parameter and susceptibility critical exponents and the functions $X_m$ and $X_{\chi}$ are scaling functions. Analogously, the reduced correlation lengths $\xi_s/L$ and $\xi_{\tau}/L_{\tau}$ take the scaling forms 
\begin{align}
\xi_s/L & = X_{\xi_s}(rL^{1/\nu},L_{\tau}/L^z), \label{corrtScaling}\\
\xi_{\tau}/L_{\tau} & = X_{\xi_{\tau}}(rL^{1/\nu},L_{\tau}/L^z). \label{corrsScaling}
\end{align}
We can also establish information about the compressibility $\kappa$ and superfluid density $\rho_s$ of the system. Under the quantum-to-classical mapping, the compressibility $\kappa = \partial \langle n\rangle/\partial \mu$  and superfluid density $\rho_s$ map, respectively, onto the spinwave stiffnesses in imaginary-time and space dimensions as
\begin{align}
\label{eqn:stifft}
\rho_{cl,\tau} & = L_{\tau}^2(\partial^2 f/\partial \theta^2)_{\theta = 0}\\
\rho_{cl,s} & = L^2(\partial^2 f/\partial \theta^2)_{\theta = 0}
\label{eqn:stiffs}
\end{align}
where $f$ is the free energy density for twisted boundary conditions (i.e. the XY spins of the classical model $\mathbf{S}_{i,\tau}$ at $\tau = 0$ ($i=0$) are at an angle $\theta$ with respect to the spins at the boundary $\tau = L_{\tau}$ ($i = L$)). Explicitly, for the XY model considered here (\ref{eqn:stifft}) takes the form\cite{TeitelJayaprakash83}
\begin{multline}
\rho_{cl,\tau} = \frac{1}{N}\sum_{i,\tau}\langle \mathbf{S}_{i,\tau} \cdot \mathbf{S}_{i,\tau+1} \rangle \\ - \frac{\beta}{N} \bigg\langle\bigg\{ {\sum_{i,\tau}\hat{k}\cdot(\mathbf{S}_{i,\tau}\times\mathbf{S}_{i,\tau +1})}\bigg\}^2\bigg\rangle
\end{multline} 
where $\hat{k}$ represents the unit vector perpendicular to the XY plane of the spins. The space stiffness $\rho_{cl,s}$ takes an analogous form. These quantities are expected to exhibit power-law scaling behavior according to the scaling forms
\begin{align}
\rho_{cl,s} & = L^{-y_s}X_{\rho_s}(rL^{1/\nu},L_{\tau}/L^z) \label{stiffsscaling} \\
\rho_{cl,\tau} & = L^{-y_{\tau}}X_{\rho_{\tau}}(rL^{1/\nu},L_{\tau}/L^z) \label{stifftscaling}
\end{align}
where $X_{\rho_s}$ and $X_{\rho_{\tau}}$ are scaling functions, while $y_s = d + z - 2$ and $y_{\tau} = d - z$ are the scale dimensions of the spinwave stiffnesses in space and imaginary-time, respectively \cite{WeichmanMukhopadhyay08}. Both stiffnesses are expected to be nonzero in the superfluid phase. In both the Mott insulator and the Mott glass phases, they are expected to vanish. (Note that the Mott glass is an \emph{incompressible} insulator.)

\section{Monte Carlo Simulations}
\label{sec:MC}
\subsection{Overview}
\label{sec:MCoverview}
Our investigation consists of Monte Carlo simulations of the classical XY model (\ref{Hc}) with both the standard single-spin-flip Metropolis\cite{MRRT53} algorithm, as well as the cluster-update Wolff\cite{Wolff89} algorithm. Both algorithms are used throughout the simulations and one ``full sweep" is defined as a Metropolis sweep over the entire lattice and a Wolff sweep. A Wolff sweep in our simulations flips a number of clusters such that the total number of spins flipped in the clusters is equal to the number of spins in the system. While the Wolff algorithm alone is sufficient to equilibrate clean systems, highly dilute systems can exhibit small disconnected clusters that the Metropolis algorithm can more effectively equilibrate. 

We simulate a range of dilutions $p = 0, 1/3, 1/2, 3/5$ and $p = p_c \approx 0.688392$ with system sizes up to $L=80$ in the space dimensions and $L_{\tau}=320$ in the imaginary-time dimension. All data need to be averaged over a large number of independent dilution configurations. This increases the computational effort needed for meaningful results. Best performance can be achieved with a rather small number of measurements sweeps, $N_m$, but a large number of disorder realizations (samples), $N_s$.\cite{BFMM98,BFMM98b} To this end, we have chosen $N_m = 500$ and $N_s = 4000-20000$ (depending on system size). To eliminate biases due to the short measurements, we use improved estimators\cite{ZWNHV15}. To ensure complete equilibration of the system we have chosen $N_{eq} = 100$ equilibration sweeps to be carried out before each measurement. We have confirmed that 100 sweeps are sufficient by comparing the results of the simulations with hot starts (spins initially randomly oriented) and cold starts (spins initially aligned) and verifying that they agree within their error bars. 
\begin{figure}
\includegraphics[width=\columnwidth]{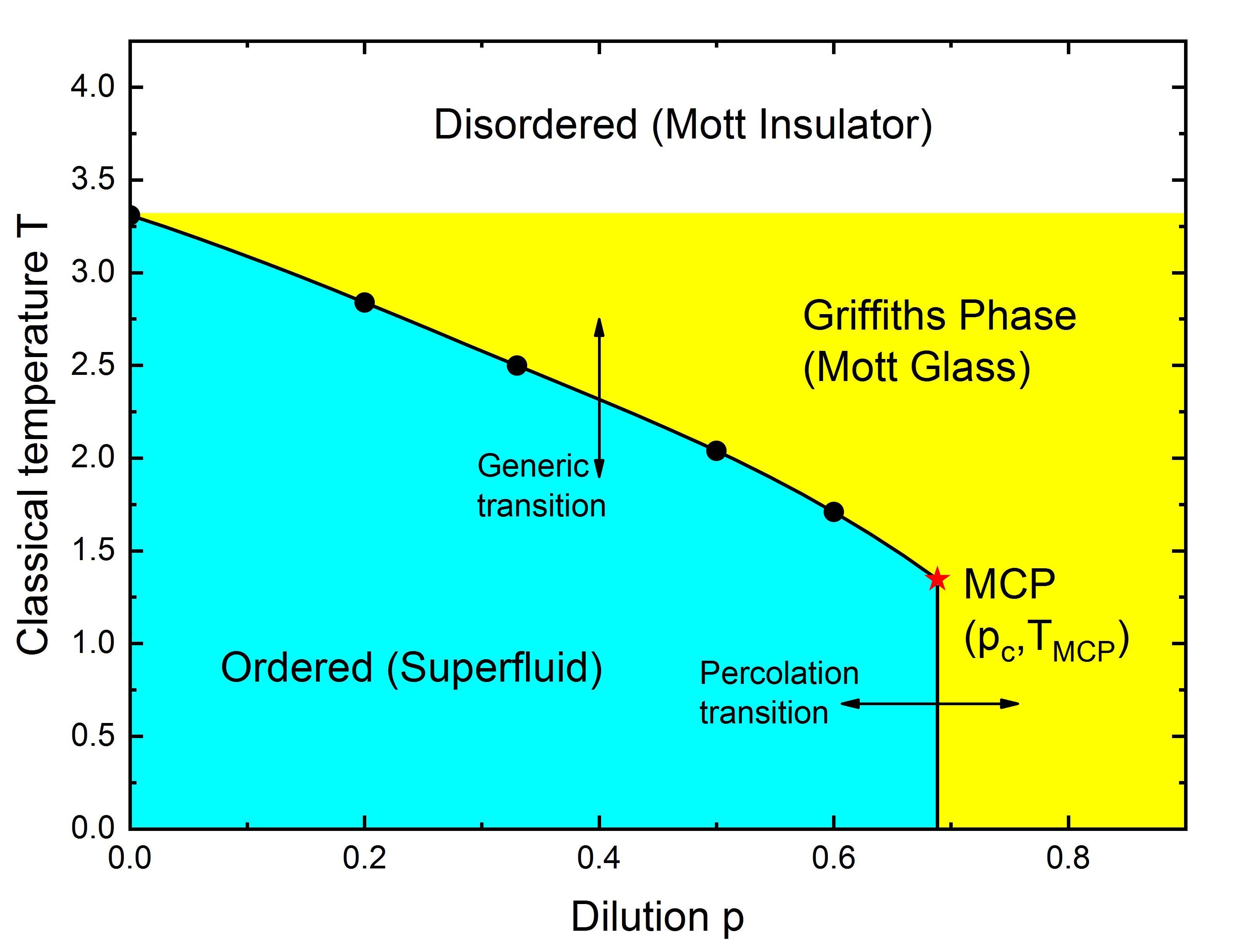}
\caption{(Color online) Phase diagram of the classical (3+1)-dimensional XY model with respect to classical temperature $T$ and dilution $p$. The multi-critical point (MCP) is estimated as the intersection of a spline interpolation of the numerical critical temperatures (dots) and the percolation transition at $p_c$. The errors of the calculated $T_c$ are smaller than the symbol size.}
\label{fig:pd}
\vspace{-10pt}
\end{figure}

The phase diagram resulting from the simulations is presented in Figure \ref{fig:pd}. As expected, the transition temperatures $T_c(p)$ decrease with increasing dilution from the clean value $T_c(0)$. The generic transition ends at the multi-critical point, which we have estimated from the intersection of a spline fit of the calculated $T_c(p)$ and the lattice percolation threshold $p_c = 0.688392$.

\subsection{Clean Critical Behavior}
First, we analyze the phase transition of the clean, undiluted system ($p=0$). Since the clean system is isotropic, we choose samples with $L = L_{\tau}$ between $10$ and $80$. The critical temperature is determined from the crossings of the $g_{av}$ vs $T$ curves for different $L$ and the corresponding crossings of the $\xi/L$ vs $T$ curves. Extrapolating to $L \rightarrow \infty$ yields a critical temperature $T_c(0) = 3.31445(3)$.
\begin{figure}
	\includegraphics[width=\linewidth]{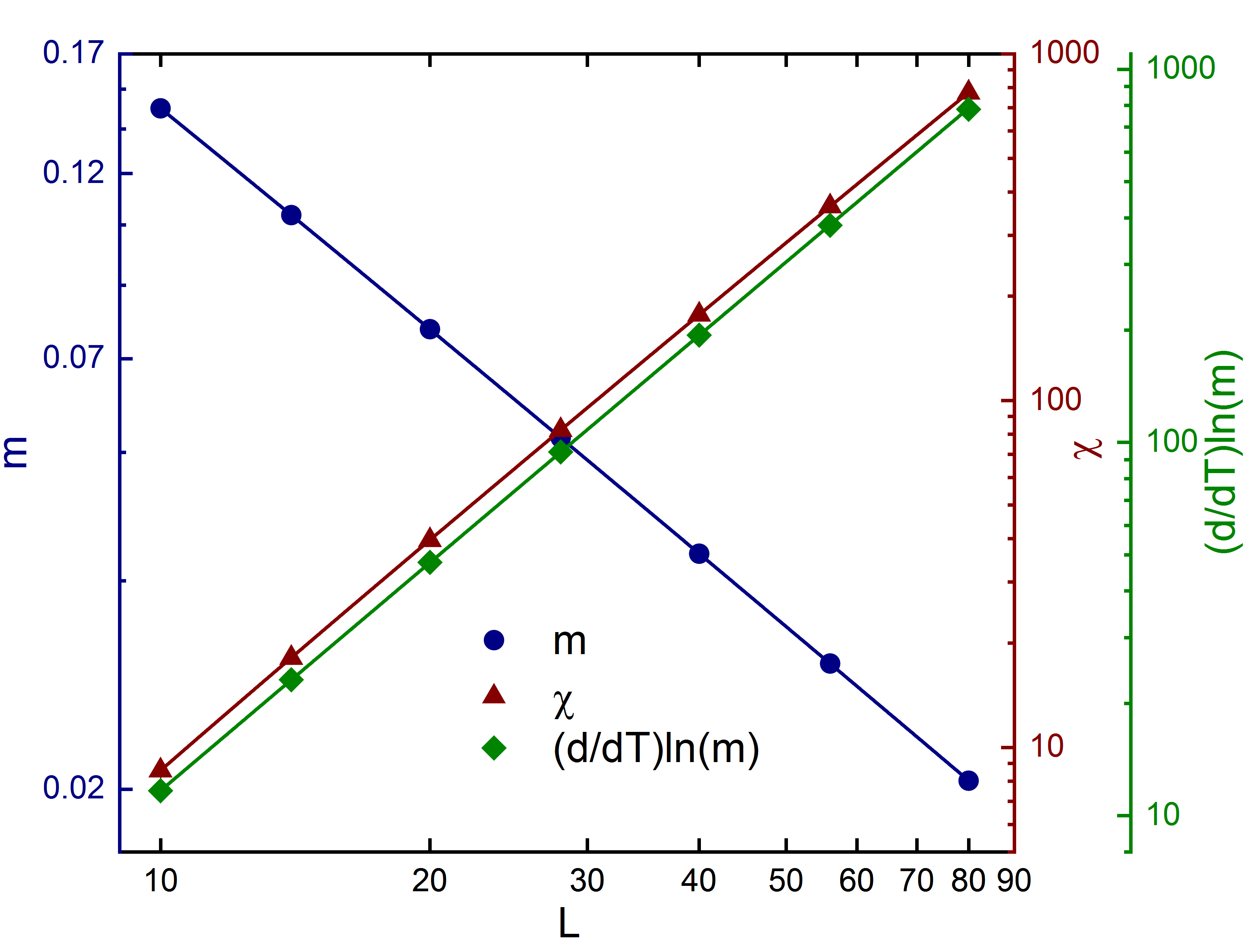}
	\caption{Order parameter $m$ and susceptibility $\chi$ vs system size $L$ for the clean case ($p = 0$). Solid lines are fits to $m = aL^{-\beta/\nu}(\ln (L/L_0))^{\omega}$ and $\chi = aL^{\gamma/\nu}(\ln (L/L_0))^{\omega}$ that yield $\beta/\nu = 1.008(12)$ and $\gamma/\nu = 2.00(1)$, respectively. Statistical errors are of the order of the symbol size.}
	\label{fig:clean_cross}  
\end{figure}

Figure \ref{fig:clean_cross} shows both order parameter and susceptibility as functions of system size right at the critical temperature. Fits of the order parameter data to the scaling form $m = aL^{-\beta/\nu}(\ln (L/L_0))^{\omega}$ are of good quality (reduced chi-squared $\tilde{\chi}^2 \approx 0.3$) and give critical exponents $\beta/\nu = 1.008(12)$ and $\omega = 0.25(8)$. Considering the same fits for various temperatures within the error bars of our critical temperature estimate, leads to variation in $\beta/\nu$ of around $0.02$. Our final estimate for the order parameter exponent is $\beta/\nu = 1.00(2)$.

Fits of the susceptibility to the scaling form $\chi = aL^{\gamma/\nu}(\ln (L/L_0))^{\omega'}$ are less stable. We fit the data to the scaling form with the irrelevant exponent fixed at it's predicted value $\omega' = 1/2$ from equation (\ref{cleanchiscaling}). This yields a critical exponent $\gamma/\nu = 2.00(2)$ with reduced chi-squared $\tilde{\chi}^2 \approx 0.65$. Susceptibility fits are more sensitive to errors in critical temperature, having a variation in $\gamma/\nu$ of about $0.04$ for temperatures within our error bar estimates. Our final estimate for the susceptibility critical exponent is $\gamma/\nu = 2.00(6)$. 

Lastly, we find the correlation length critical exponent via slopes of the Binder cumulant $g_{av}$, reduced correlation length $\xi/L$ and logarithm of the order parameter $\ln(m)$, with respect to temperature. Equation (\ref{cleandlnmdtscaling}) predicts a value of $\nu = 1/2$ for the correlation length critical exponent for $(d/dT)\ln(m)$ and universality implies the same scaling form holds for $g_{av}$ \& $\xi/L$. Fitting the data for $(d/dT)\ln(m)$ to the scaling form $aL^{1/\nu}\ln(L/L_0)^{\bar{\omega}}$ with irrelevant exponent fixed at the theoretical value $\bar{\omega} = 1/10$ yields the critical exponent $\nu = 0.50(2)$ for an acceptable fit ($\tilde{\chi}^2 \approx 4$). Similar analysis for $(d/dT)g_{av}$ and $(d/dT)(\xi/L)$ yields $\nu = 0.50(2)$ and $\nu = 0.49(4)$, respectively. Our final estimate for the correlation length critical exponent is $\nu = 0.50(6)$.

Finally, we note that pure power-law fits to the data show significantly larger $\tilde{\chi}^2$ values. This further justifies the logarithmic corrections in the scaling forms (\ref{cleanmscaling}) - (\ref{cleandlnmdtscaling}). In summary, all of our Monte Carlo results for the clean case are in good agreement with the scaling theory of Ref. \onlinecite{Kenna04}.

\subsection{Disordered Case: Generic Transition}
\label{sec:GenericTrans}
The finite-size scaling analysis of the generic transition, ($0<p<p_c$) is carried out as described in Sec. \ref{sec:AFSS}. Determining a full set of critical exponents requires first finding the optimal shapes and calculating the dynamical exponent $z$ in order to fix the second argument of our scaling forms (\ref{BinderPowScaling}) - (\ref{stifftscaling}). This is achieved using the iterative procedure also outlined in Sec. \ref{sec:AFSS}.

\begin{figure}
	\includegraphics[width=\columnwidth]{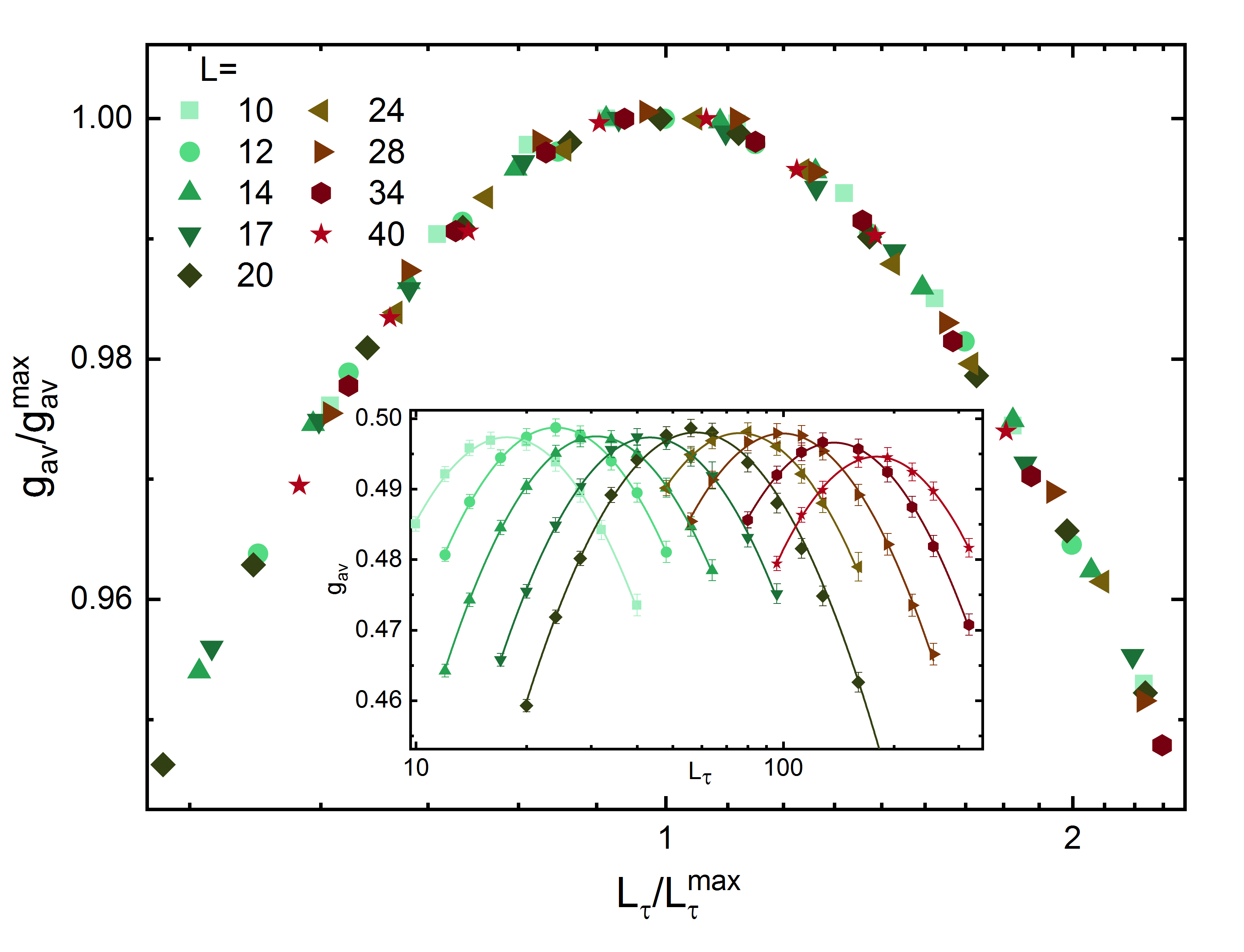}
	\caption{Binder cumulant $g_{av}$ as a function of $L_{\tau}$ for several $L$ and dilution $p=0.5$ at the critical temperature $T_c = 2.037$. Plotting $g_{av}/g_{av}^{\text{max}}$ in the main panel eliminates the leading additive correction to scaling from the analysis.}
	\label{fig:domescaling}
\end{figure}

\begin{figure}
	\includegraphics[width=\columnwidth]{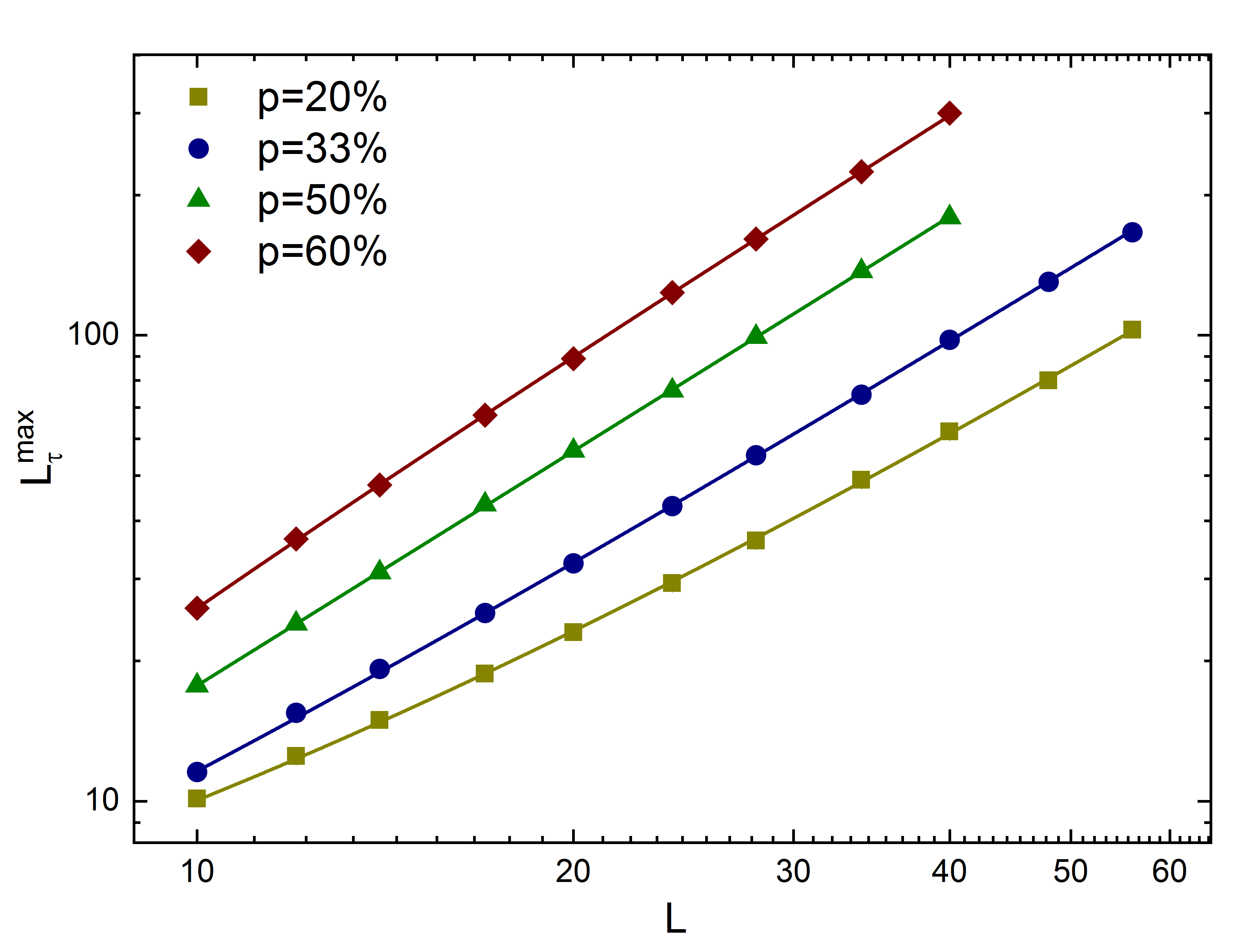}
	\caption{Log-log plots of $L_{\tau}^{\text{max}}$ vs $L$. Solid lines are fits to $L_{\tau}^{\text{max}} = aL^z(1+bL^{-\omega})$ yielding $z = 1.672(9)$ and $\omega = 1.18(5)$. Statistical errors are of the order of the symbol size.}
	\label{fig:zplot}
\end{figure}

Figures \ref{fig:domescaling} and \ref{fig:zplot} show an example of this analysis. Specifically, Fig. \ref{fig:domescaling} presents the Binder cumulant $g_{av}$ for the dilution $p=0.5$ as a function of $L_{\tau}$ for system sizes $L = 10 - 40$ at the estimated critical temperature. The raw data are shown in the inset; as expected, $g_{av}^{\text{max}}$ at the critical point is (roughly) independent of $L$ and exhibits a maximum at $L_{\tau}^{\text{max}}$ for each system size. Remaining variation of $g_{av}^{\text{max}}$ is due to the uncertainty in $T_c$ for the large system sizes and corrections to scaling for small system sizes (both of these will be discussed further below). The main panel is a scaling plot demonstrating that the Binder cumulant fulfills the scaling form (\ref{BinderPowScaling}) to a high degree of accuracy and variations due to uncertainty in $T_c$ and corrections to scaling simply shift the $g_{\text{av}}$ vs. $L_{\tau}$ curves up or down. Corresponding scaling plots were also constructed with analogous results for the remaining dilutions $p=1/5,1/3,$ and $3/5$. 

Determining $z$ requires analyzing the position $L_{\tau}^{\text{max}}$ of these maxima which we have found via quadratic fits of $g_{av}$ vs $\ln L_{\tau}$. Plots of $L_{\tau}^{\text{max}}$ vs $L$ are shown in Figure \ref{fig:zplot}. As can be seen, the data show significant corrections to scaling (deviations from straight lines), especially for smaller dilutions. Neglecting them by fitting the data via pure power laws would yield only effective, scale-dependent exponents. Therefore, we include the leading-order correction to scaling via the ansatz $L_{\tau}^{\text{max}} = aL^z(1+bL^{-\omega})$ with dilution-independent critical exponents $z$ and $\omega$ but dilution-dependent prefactors $a$ and $b$. This yields true asymptotic, scale-independent critical exponents. Combined fits of all four dilution data sets gives exponents $z = 1.672(9)$ and $\omega = 1.18(5)$ with an acceptable reduced chi-squared $\tilde{\chi}^2 \approx 2.69$. If we consider the robustness of the combined fits against removal of upper and lower data points from each set as well as removal of entire dilution sets, we come to an estimate for the dynamical critical exponent of $z = 1.67(4)$. We also note that the leading corrections to scaling vanish close to $p = 1/2$ where the prefactor $b$ changes sign and is effectively zero for our fits of $p = 1/2$. The vanishing of these corrections is also reinforced by the comparison of pure power-law fits and fits to scaling forms including subleading corrections. For the $p=1/2$, power-law fits yield $z = 1.671(3)$, where the fits including the subleading corrections yields $z = 1.66(1)$. The global, dilution independent value for the dynamical exponent is also bracketed nicely by the values obtained upon pure power-law fits of the largest system sizes for the dilutions $p = 1/3$ and $p = 3/5$, which yield $z = 1.592(6)$ and $z = 1.767(7)$, respectively. To estimate the error of $z$ stemming from the uncertainty in $T_c$, we have repeated the analysis for appropriately chosen temperatures slightly above and below our estimate for $T_c$. Variation of the dynamical exponent within this range of temperatures is about $0.03$. After considering this uncertainty in $T_c$, statistical error and the robustness of our fits, we come to our final estimate of the dynamical exponent $z = 1.67(6)$. 

\begin{figure}
	\includegraphics[width=\columnwidth]{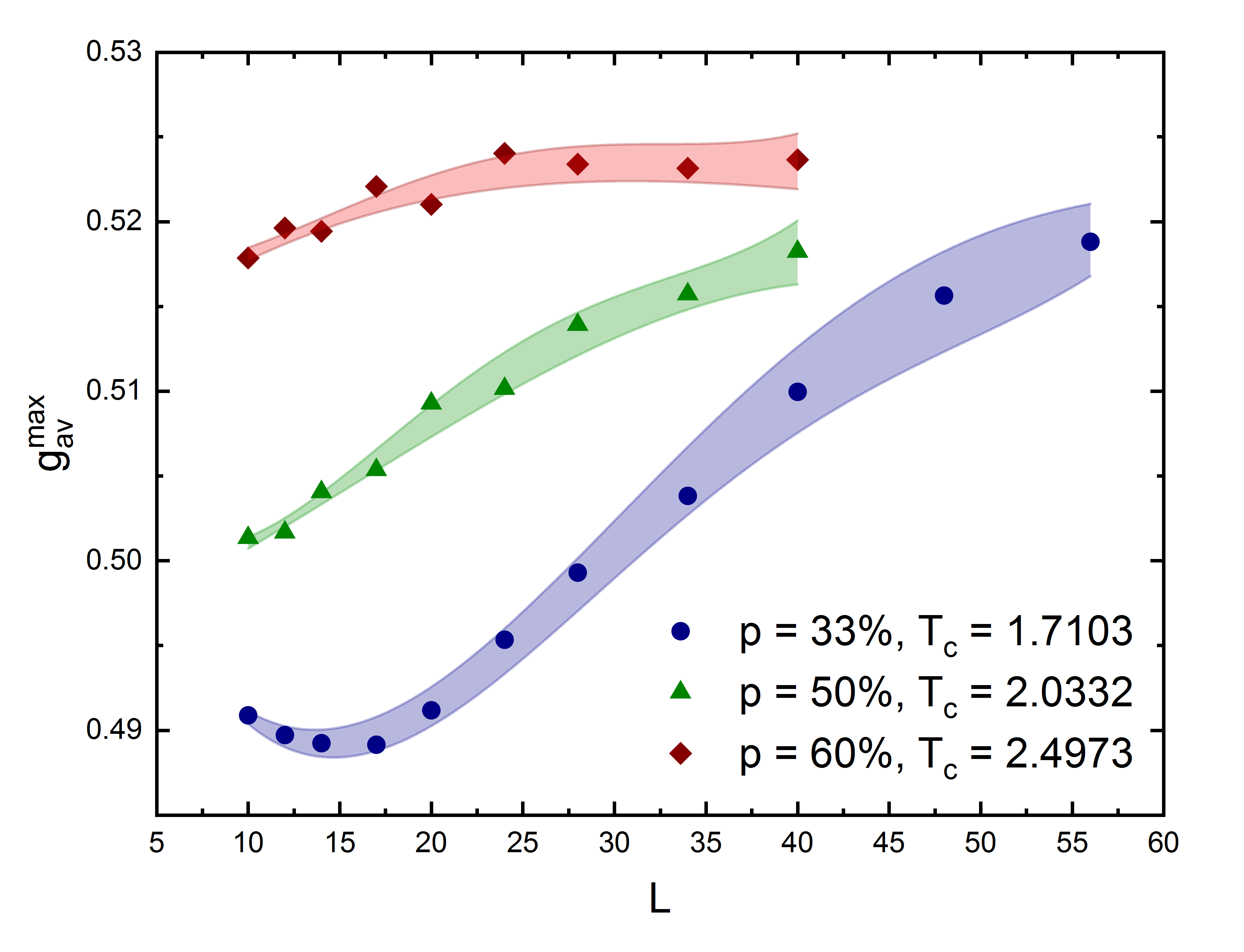}
	\caption{$g_{av}^{\text{max}}$ vs $L$ for the improved estimates for $T_c$. Shaded regions represent the the range values for which the criterion is also satisfied. From this we estimate an error on $T_c$ of no more than $0.0003$. Statistical errors are of the order of the symbol size or smaller. The remaining variation of $g_{av}^{\text{max}}$ likely stems from the discreteness of $L_{\tau}$.}
	\label{fig:binder-converge}
\end{figure}

To complete our set of critical exponents, we now analyze the Monte Carlo runs for systems of optimal shape and in the vicinity of their critical temperature $T_c(p)$. With $L_{\tau}/L^z$ fixed by the optimal shapes found above, the scaling forms (\ref{mScaling}) and (\ref{chiScaling}) are then used to extract $\beta/\nu$ and $\gamma/\nu$ from the $L$ dependence of the order parameter $m$ and susceptibility $\chi$ at $T_c(p)$. We again fit the data with leading corrections to scaling included via the ansatz $m = aL^{-\beta/\nu}(1+bL^{-\omega})$ and $\chi = aL^{\gamma/\nu}(1+bL^{-\omega})$ with universal exponents but dilution-dependent prefactors. However, the combined fits of these data proved to be very sensitive to small changes in $T_c(p)$ (much more so than the fit determining $z$). This indicates that our critical temperature estimates (originally found from the crossings of the curves of dimensionless quantities versus temperature) are not the true critical temperatures. Thus, to improve our critical temperature estimates, we impart the criterion that at criticality the value of $g_{av}^{\text{max}}$ should approach a dilution-independent value as $L \longrightarrow \infty$. We can adjust our estimates for $T_c(p)$ until this criterion is satisfied, with $g_{av}^{\text{max}}$ approaching dilution and system size independent values, as is shown in Figure \ref{fig:binder-converge}.\cite{Vojtaetal16} This adjustment of the critical temperatures yields our final estimates: $T_c(1/5) = 2.837$, $T_c(1/3) = 2.4973$, $T_c(1/2) = 2.0332$, $T_c(3/5) = 1.7103$. The data can also be seen to satisfy this criterion for a small range of temperatures, thus we assign an error to our estimated critical temperatures of no more than $\pm 0.0003$. The data in Figure \ref{fig:binder-converge} clearly demonstrates that the systems with dilutions $p=1/3$ and $p=1/2$ show pronounced corrections to scaling. They are still crossing over from the clean critical fixed point to the asymptotic regime even at the largest $L$. Moreover, $g_{av}^{\text{max}}$ for small system sizes exhibits non-monotonous behavior, from which we conclude that there are at least two corrections to scaling contributing for the smallest dilutions and system sizes.

\begin{figure}
	\includegraphics[width=\columnwidth]{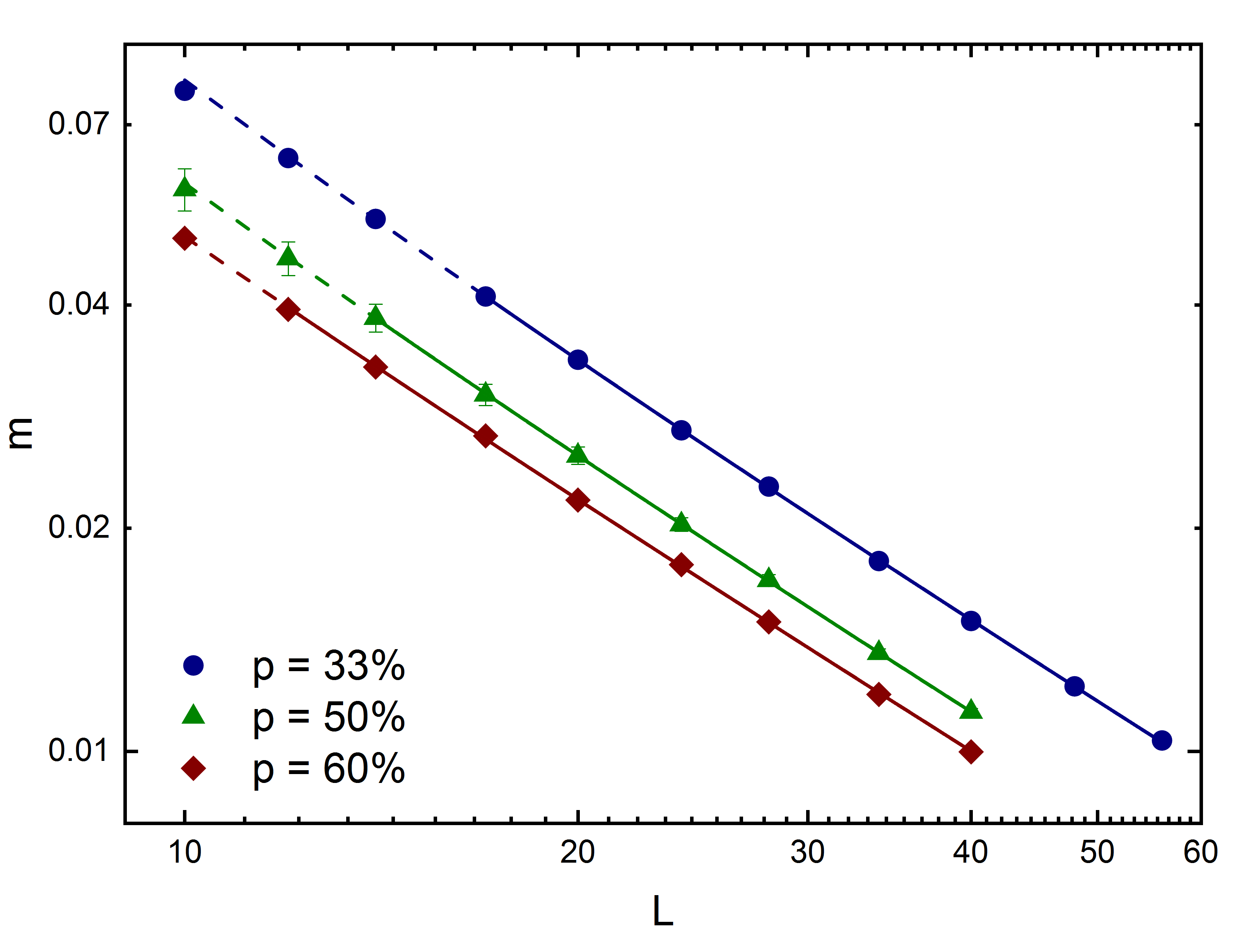}
	\caption{Log-log plot of $m$ vs $L$ at the critical temperature. Solid lines are fits to $m = aL^{-\beta/\nu}(1+bL^{-\omega})$ that yield $\beta/\nu = 1.087(11)$ and $\omega = 1.22(7)$. Lines are dashed in regions that are not included in the fit. Statistical errors are of the order of the symbol size unless shown explicitly in the plot.}
	\label{fig:mag_L}
\end{figure}

With the improved estimates for $T_c$, we proceed to fit the three largest dilutions ($p = 1/5, 1/3, 3/5$) with the above scaling ansatz to find $\beta/\nu$ and $\gamma/\nu$. Order parameter $m$ versus system size $L$ for the three dilutions is shown in Figure \ref{fig:mag_L}. We perform a combined fit with $m = aL^{-\beta/\nu}(1+bL^{-\omega})$. Leaving out the system sizes most affected by the second sub-leading corrections to scaling mentioned above, we get good fits ($\tilde{\chi}^2 \approx 0.43$) that result in a critical exponent $\beta/\nu = 1.087(11)$ and correction exponent $\omega = 1.22(7)$. Fits to the same data for slightly adjusted temperatures within the estimated error ($T_c \pm 0.0003$) lead to variation in the critical exponent of about 0.02. Our final estimate for the order parameter critical exponent then reads $\beta/\nu = 1.09(3)$.

\begin{figure}
	\includegraphics[width=\columnwidth]{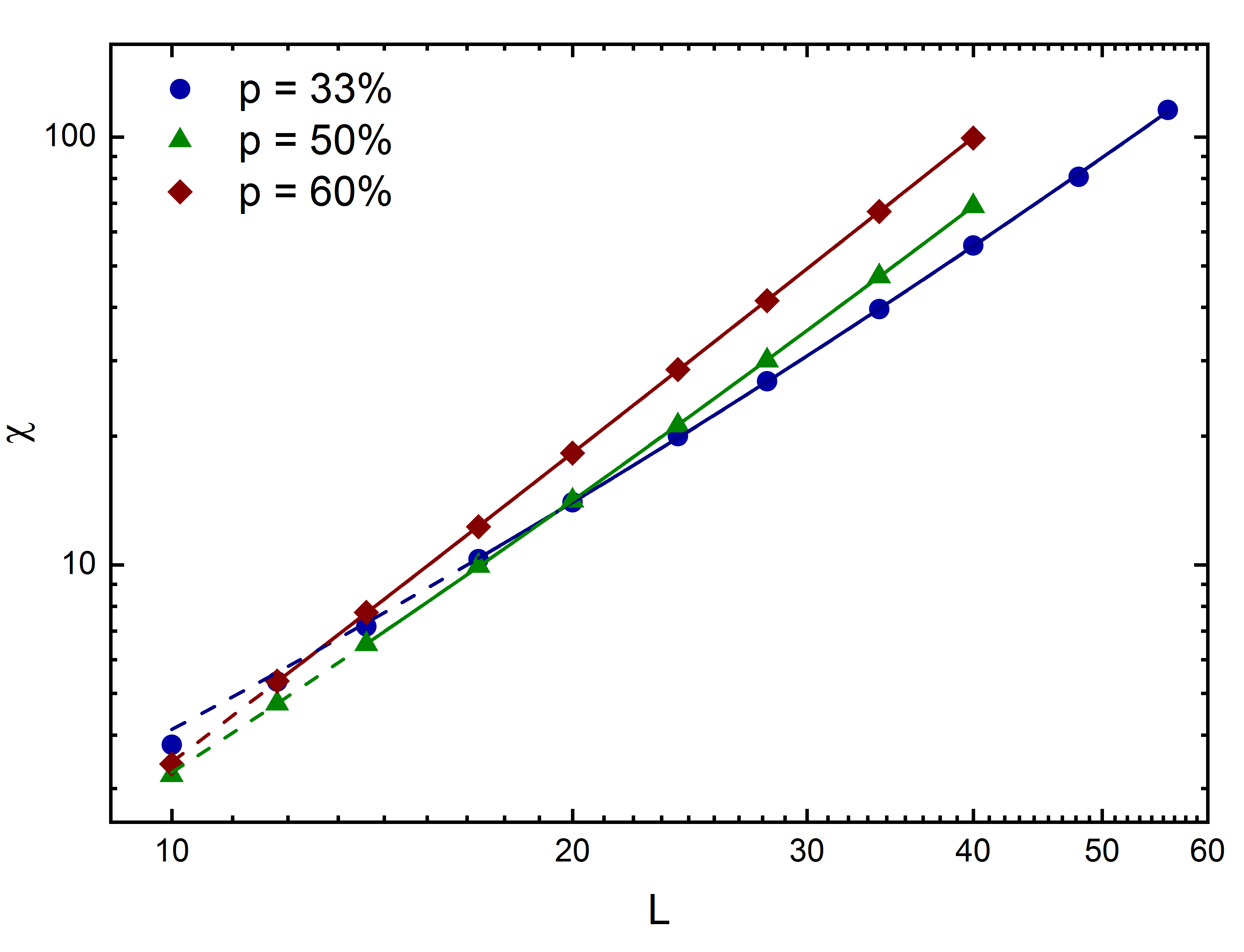}
	\caption{Log-log plot of $\chi$ vs $L$ at the critical temperature. Solid lines are fits to $\chi = aL^{\gamma/\nu}(1+bL^{-\omega})$ that yield $\gamma/\nu = 2.495(7)$ and $\omega = 1.16(2)$. Lines are dashed in regions that are not included in the fit. Statistical errors are of the order of the symbol size.}
	\label{fig:susc_L}
\end{figure}

Figure \ref{fig:susc_L} shows the order parameter susceptibility $\chi$ as a function of system size $L$ at criticality. Fitting to the ansatz with leading-order corrections $\chi = aL^{\gamma/\nu}(1+bL^{-\omega})$, and again dropping the system sizes most affected by sub-leading order corrections, we arrive at a good fit ($\tilde{\chi}^2 \approx 1.3$) that yields the critical exponent $\gamma/\nu = 2.495(7)$ and correction exponent $\omega = 1.16(2)$. After considering the uncertainties in $T_c$ and fit range, as we did for $\beta/\nu$, we come to the final estimate for the susceptibility exponent $\gamma/\nu = 2.50(3)$.

\begin{figure}
	\includegraphics[width=\columnwidth]{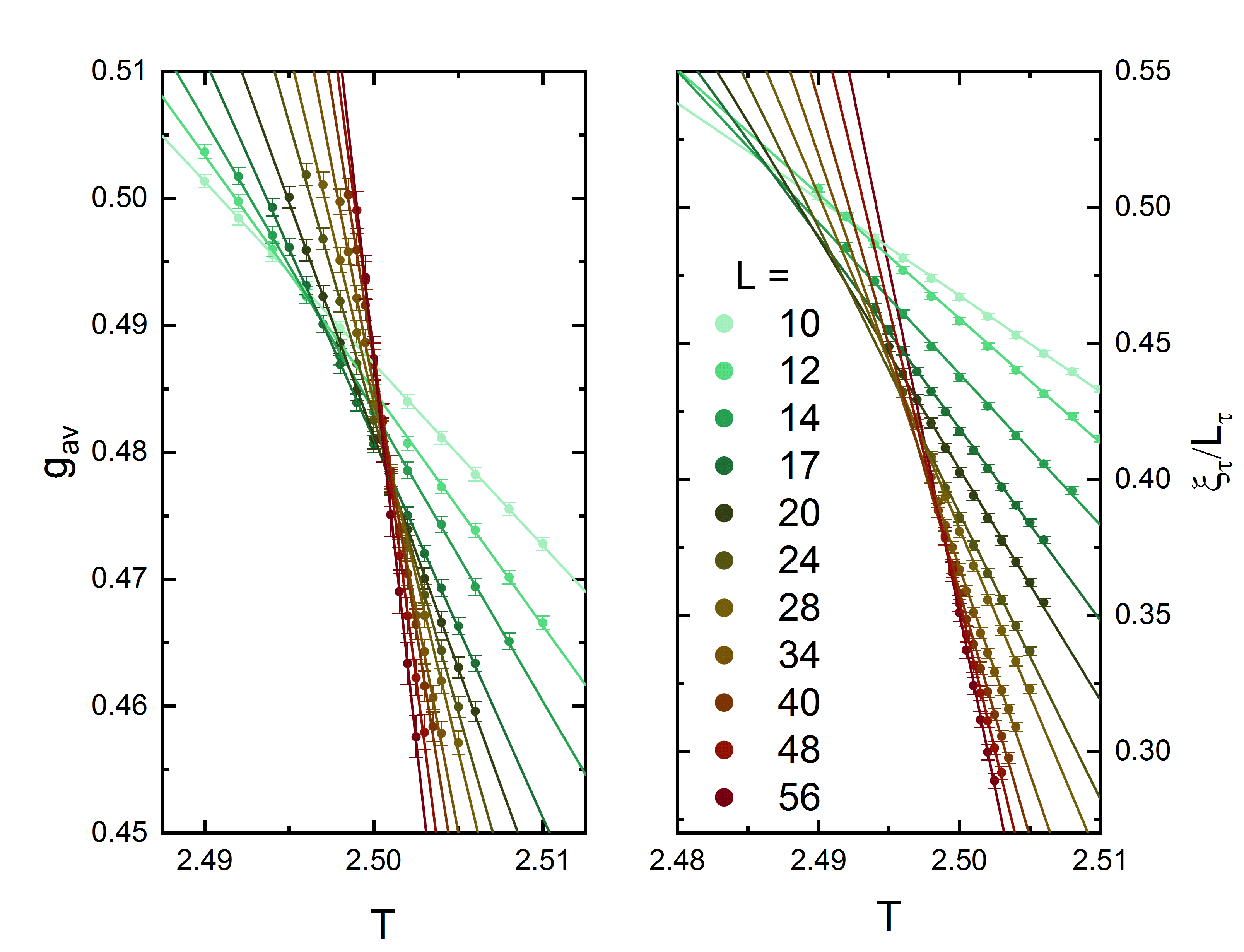}
	\caption{Binder cumulant $g_{av}$ and reduced correlation function $\xi_{\tau}/L_{\tau}$ for systems of optimal shape and dilution $p =1/3$. Plotted are system sizes $L = 10 - 56$ with increasing slope.}
	\label{fig:gav_corr_cross}
\end{figure}

\begin{figure}
	\includegraphics[width=\columnwidth]{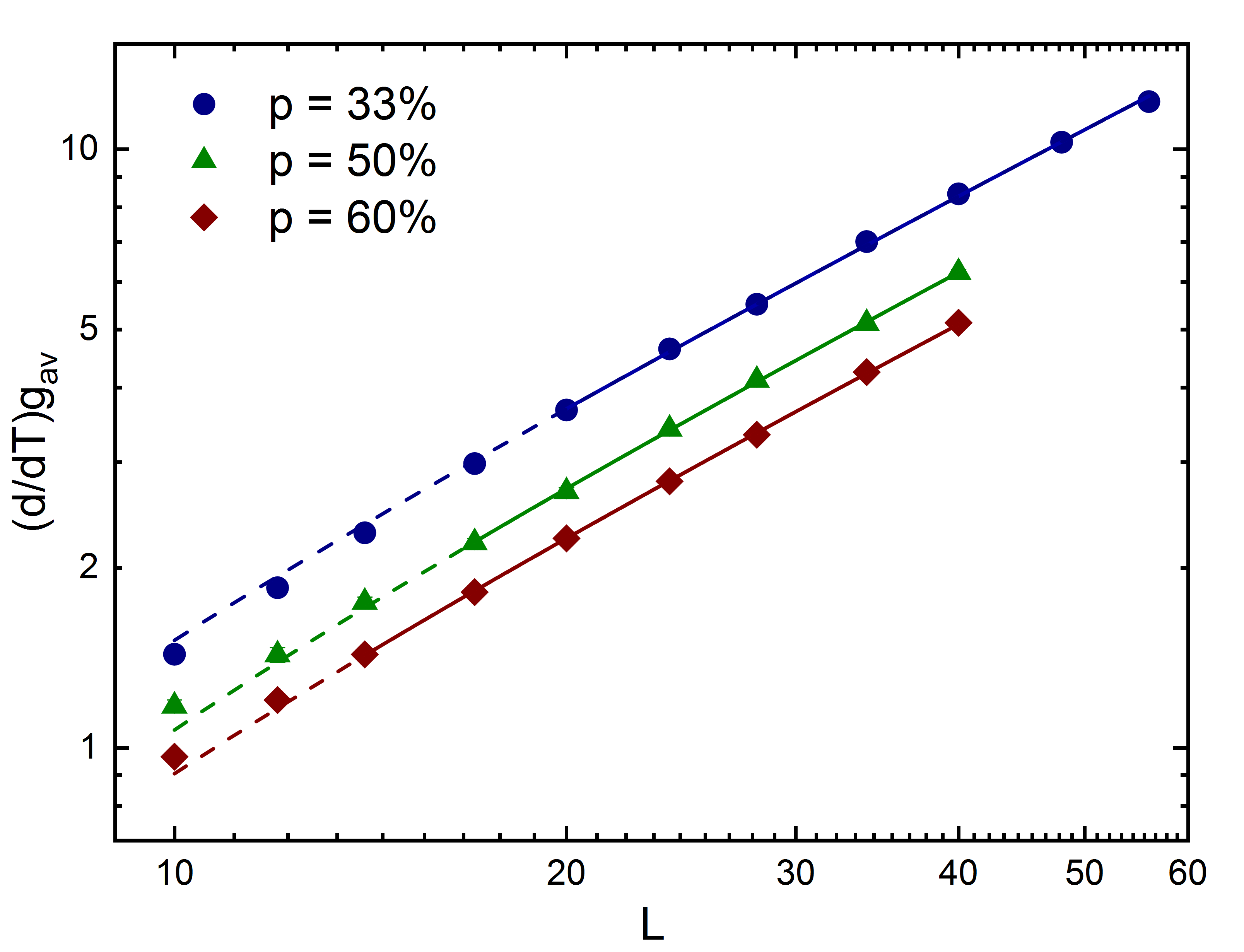}
	\caption{Log-log plot of $(d/dT)g_{av}$ vs $L$. Solid lines are fits to $g_{av} = aL^{1/\nu}(1+bL^{-\omega})$ that yield $\nu = 0.90(2)$ and $\omega = 1.17(8)$. Lines are dashed in regions that are not included in the fit. Statistical errors are of order of the symbol size.}
	\label{fig:dTgav_L}
\end{figure}

We now move to determining the correlation length critical exponent. This can be determined by considering the slopes of $g_{av}$ and $\xi_{\tau}/L_{\tau}$ as functions of temperature. Figure \ref{fig:gav_corr_cross} shows off-critical data $g_{av}$ and $\xi_{\tau}/L_{\tau}$ for dilution $p=1/3$, as functions of temperature. Since both quantities have scale dimension zero, they should cross directly at the critical temperature. However, it is clear in the data that a shift occurs in these crossings for increasing system size $L$, thus we still expect significant corrections to scaling. Equations (\ref{BinderPowScaling}) and (\ref{corrtScaling}) show that the correlation exponent can be extracted from finite-size scaling of $(d/dT)g_{av}$ and $(d/dT)\xi_{\tau}/L_{\tau}$, which each vary as $L^{1/\nu}$ with system size. Extracting the slopes of each of these functions is done by linear fits to the data in the vicinity of the critical temperature. Figure \ref{fig:dTgav_L} shows the slopes of the Binder cumulant $g_{av}$ as a function of system size. Again, to account for the corrections to scaling, we fit this data with the ansatz scaling form $aL^{1/\nu}(1+bL^{-\omega})$. Combined fits to $(d/dT)g_{av}$ lead to $\nu = 0.90(2)$ and $\omega = 1.17(8)$ with a reduced chi-squared $\tilde{\chi} \approx 2.2$. Similar fits of the reduced correlation length $\xi_{\tau}/L_{\tau}$ are of good quality ($\tilde{\chi}^2 \approx 1.15$) when the smallest system sizes are left out giving a correlation exponent of $\nu = 0.894(4)$ and correction exponent $\omega = 1.16(10)$. Similar analysis carried out on $(d/dT)\xi_{s}/L$ yields nearly identical results. Considering the robustness of the fits against removal of upper and lower data points, we are led to a somewhat larger error, leading to a final estimate that reads $\nu = 0.90(5)$. 

The critical exponents must satisfy the hyperscaling relationship $2\beta/\nu + \gamma/\nu = d + z$, where $d=3$ is the spatial dimension. Our values $\beta/\nu = 1.09(3)$, $\gamma/\nu = 2.50(3)$, and $z = 1.67(6)$ fulfill this relationship nicely within the error bars. We can also assign a value to the anomalous dimension $\eta$, defined via the decay of the critical correlation function in space, $G(\mathbf{x}) \sim |\mathbf{x}|^{-(d+z-2+\eta)}$. It measures the deviation of $G$ from a hypothetical Gaussian theory \footnote{A purely Gaussian theory would predict a correlation function that decays as $G \sim |\mathbf{x}|^{-(d+z)+2}$ with $z$ the dynamical exponent of the system. The anomalous dimension is the deviation of the exponent from this power-law behavior.}. This anomalous dimension $\eta$ can be calculated via the relationship $\eta = 2 - \gamma/\nu$, giving the result $\eta = -0.50(3)$.  Additionally, the inequality $d\nu > 2$ is now fulfilled for our correlation exponent $\nu = 0.90(5)$. Because the critical exponents satisfy the hyperscaling relationship and the values of the exponent $\omega$ that governs the corrections to scaling are consistent across the range of fits, we can conclude that the critical exponent estimates that we have obtained are the true asymptotic critical exponents.

\subsection{Superfluid Density}
\label{sec:SFdensity}
\begin{figure}
	\includegraphics[width=\columnwidth]{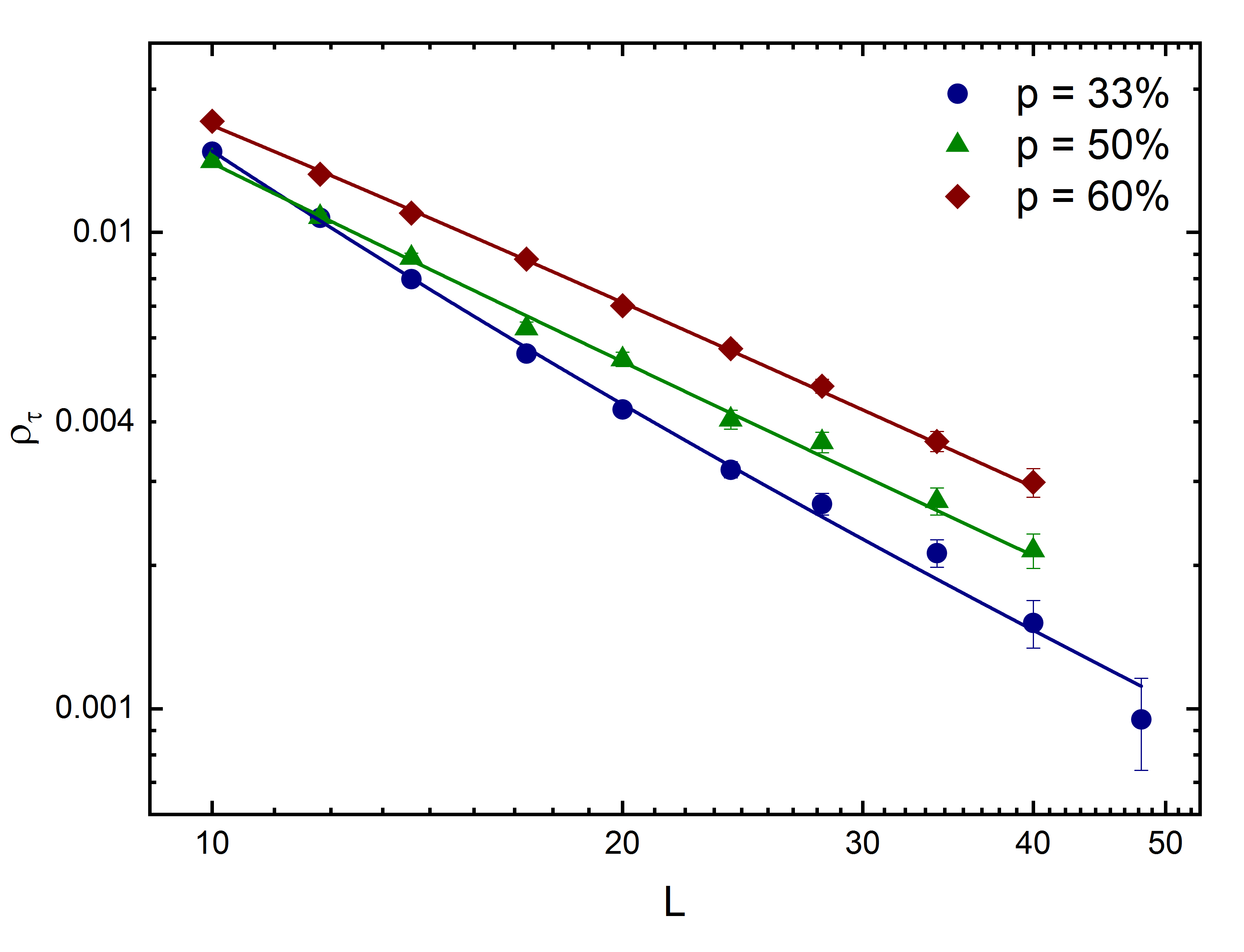}
	\caption{Log-log plot of $\rho_{\tau}$ vs $L$. Solid lines are fits to $\rho_{\tau} = aL^{-y_{\tau}}(1+bL^{-\omega})$ that yield $y_{\tau} = 1.32(2)$ and $\omega = 1.19(6)$.}
	\label{fig:stifft_L}
\end{figure}

A final result from our simulations is the critical behavior of the compressibility $\kappa$ and superfluid density $\rho_s$. This is determined by considering the behavior of the spinwave stiffness of the classical Hamiltonian (\ref{Hc}) in space and imaginary-time dimensions for optimally shaped systems right at the critical temperatures for the dilutions $p = 1/3, 1/2, 3/5$. Both observables, $\rho_{cl,s}$ and $\rho_{cl,\tau}$, are very close to zero and thus, noisy. A plot of $\rho_{cl,\tau}$ vs $L$ is shown in Figure \ref{fig:stifft_L}. Corrections to scaling are clearly relevant still, so we perform fits with first-order corrections $\rho_{cl,\tau} = aL^{-y_{\tau}}(1+bL^{-\omega})$. Good fits can be obtained over the entire data set despite the noisy large system sizes ($\tilde{\chi}^2 \approx 1.03$), yielding $y_{\tau} = 1.32(1)$ and $\omega =  1.19(6)$. The fit is surprisingly stable against removal of data points and dilution sets. We quote our final estimate of this exponent as $y_{\tau} = 1.32(2)$. This satisfies the generalized Josephson relation \cite{FWGF89} for the compressibility $y_{\tau} = d - z$ within error bars.

Spinwave stiffness in the space dimensions is much smaller and thus has larger statistical errors. Independent fits were not possible for this data set. However, we fit the data with the functional form $y_s = aL^{-y_s}(1+bL^{-\omega})$ fixing the exponents via the generalized Josephson relations $y_s = d + z - 2$. Fixing $y_s = 2.67$ and $\omega = 1.18$ (from earlier fits) yields a reasonable fit to the data ($\tilde{\chi}^2 \approx 0.03$), in agreement with expectations. 

\subsection{Percolation Transition}
\label{sec:PercTrans}
So far, we have analyzed ``generic" transitions that are driven by tuning of the (classical) temperature for dilutions $p<p_c$. Another type of transition -- the percolation transition -- can occur by tuning the dilution concentration $p$ through the percolation threshold $p_c$ of the lattice at very low temperature. The critical behavior of these transitions is entirely dependent on the critical geometry of percolating lattice with the dynamics of the rotor model unaffected, remaining locally ordered on each percolating cluster. A theory has been developed \cite{VojtaSchmalian05b} that predicts the critical behavior of this percolation quantum phase transition. These predictions give exponents $\beta = 0.417$, $\gamma = 4.02$, $\nu = 0.875$, and $z = 2.53$. Note that the static exponents $\beta$ and $\nu$ as well as the percolation threshold $p_c$ agree with the corresponding 3D classical percolation values (see, e.g., Refs. \onlinecite{WZZGD},\onlinecite{XWPD},\onlinecite{StaufferAharony_book91})

\begin{figure}
	\includegraphics[width=\linewidth]{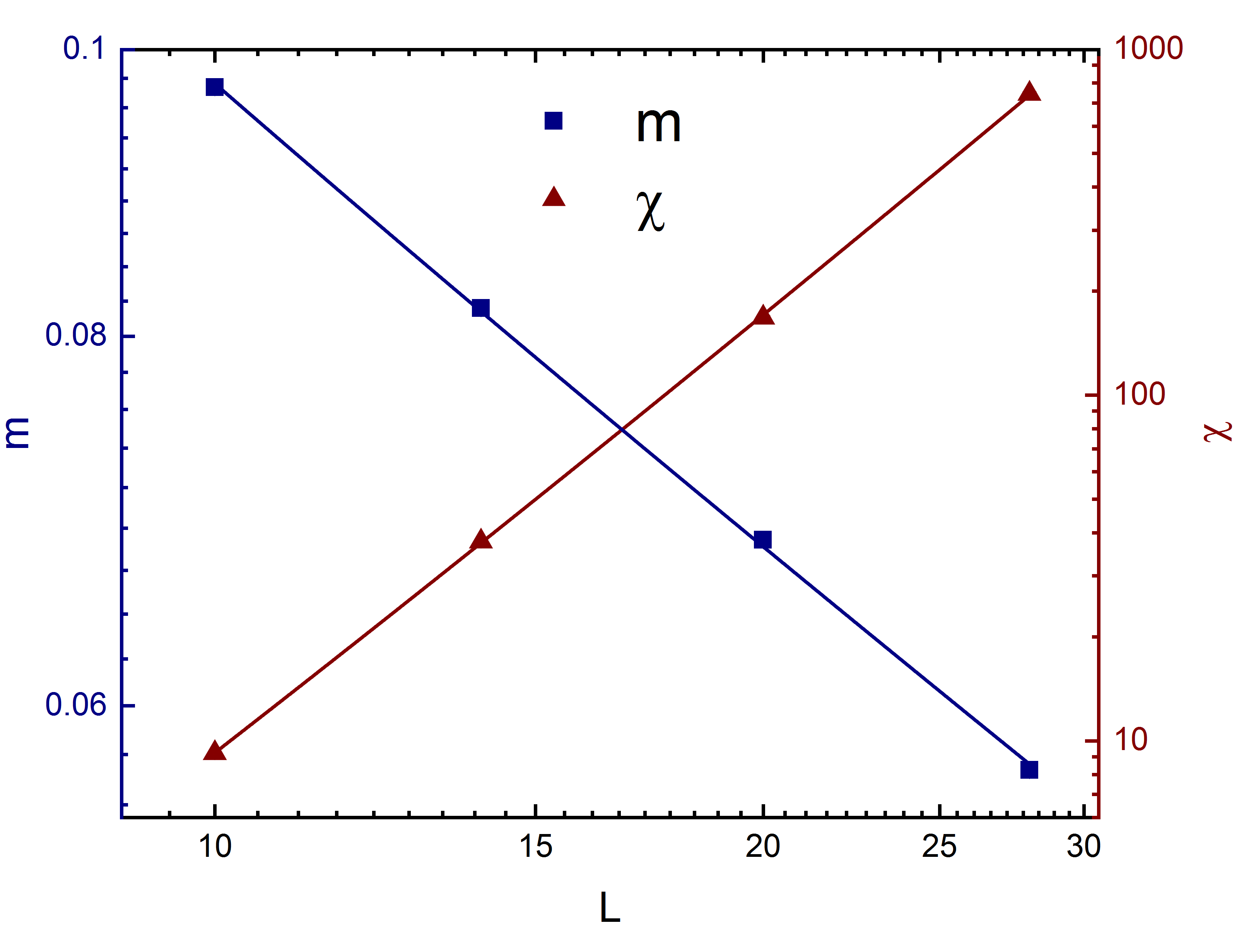}
	\caption{Log-log plot of observables $m$ and $\chi$ for the percolation transition at $p_c = 0.688392$ and $T=1.0$. Dashed lines are fits to the expectations \cite{VojtaSchmalian05b}. Statistical errors are of the order of the symbol size.}
	\label{fig:perc_L}
\end{figure}

To test these predictions we perform simulations with dilution right at the percolation threshold $p = p_c = 0.688392$ and temperature $T = 1.0$, well below the estimated multi-critical temperature $T_{MCP} \approx 1.35$. The large value of the predicted $z$ leads to the need for very large system sizes $L_{\tau}$ to confirm the dynamical critical exponent. To reduce the numerical effort, we simulated systems with the dynamical exponent fixed at its predicted value $z = 2.53$ and used these optimally shaped systems to confirm the remaining critical exponents. Figure \ref{fig:perc_L} shows both order parameter $m$ and susceptibility $\chi$ for these systems up to $L = 28$. Considering the small system sizes in our data, we fit both sets to their predicted scaling forms with first-order corrections included. For the order parameter exponent, theory predicts $\beta/\nu \approx 0.47657$. Fitting the data to the form $m = aL^{-\beta/\nu}(1+bL^{-\omega})$ with the critical exponent $\beta/\nu$ fixed at the predicted value, leads to a good fit ($\tilde{\chi}^2 \approx 1.41$) with irrelevant exponent $\omega = 0.99(12)$. Similarly, for the susceptibility exponent theory predicts $\gamma/\nu \approx 4.59429$. Fitting this data to $\chi = aL^{\gamma/\nu}(1+bL^{-\omega})$ while fixing the critical exponent $\gamma/\nu$ to it's predicted value, leads to fits of lesser quality ($\tilde{\chi}^2 \approx 5.31$), but still within reasonable agreement with the theory, and giving an irrelevant exponent $\omega = 1.26(58)$.

\section{Conclusions}
\label{sec:Conclusions}
In conclusion, we have carried out large-scale Monte Carlo simulations to determine the critical behavior of the superfluid-Mott glass quantum phase transition in three space dimensions. To do so we have mapped a site-diluted quantum rotor model with commensurate filling and off-diagonal disorder onto a (3+1)-dimensional classical XY model, and simulated it via the standard Metropolis and Wolff algorithms.
 
In the absence of disorder, the superfluid-Mott insulator transition falls into the four-dimensional XY universality class which features mean-field critical behavior with logarithmic corrections. The correlation exponent takes the value $\nu = 1/2$ that violates the Harris criterion $d\nu > 2$. As a consequence, the superfluid-Mott glass transition occurring in the disordered case shows critical behavior differing from that of the clean case.    

This superfluid-Mott glass transition exhibits a conventional finite-disorder quantum critical point with power-law dynamical scaling $\xi_{\tau} \sim \xi_s^z$ between the correlation time and length. This agrees with a general classification of disordered quantum phase transitions based on rare region dimensionality in the system \cite{Vojta06,VojtaSchmalian05b}. For the classical (mapped) Hamiltonian (\ref{Hc}), the rare regions are infinitely-long rods in the time-dimension, giving a rare region dimensionality $d_{RR} = 1$. Comparing this to the lower critical dimension of the classical XY model $D_c^- = 2$ we can see that $d_{RR} < D_c^-$, putting the system into class A (of conventional power-law scaling), as designated by the classification scheme. 

\begin{table}
	\begin{tabular}{|l|c|c|c|c|c|}
		\hline
		Our results & $z$ & $\beta/\nu$ & $\gamma/\nu$ & $\nu$ & $\eta$ \\ \hline
		Clean & $\mathit{1}$ & $1.00(2)$ & $2.00(6)$ & $0.50(5)$ & $0.00(5)$ \\ \hline
		Diluted & $1.67(6)$ & $1.09(3)$ & $2.50(3)$ & $0.90(5)$ & $-0.50(3)$ \\ \hline
		Percolation & $\mathit{2.53}$ & $\mathit{0.477}$ & $\mathit{4.594}$ &  $\mathit{0.875}$ & $\mathit{-2.594}$ \\
		\hline
	\end{tabular}
	\caption{Critical exponents found in this work. Italic values are not calculated directly but represent theoretical values that we have used and/or confirmed in the simulations.}
	\label{table1}
\end{table}
For the generic transition occurring for dilutions $p$ below the lattice percolation threshold $p_c$, we find universal, dilution-independent critical exponents from our Monte Carlo data. These exponents, summarized in table \ref{table1}, satisfy the hyperscaling relation as well as the Harris criterion. We have also considered the percolation transition that occurs across the percolation threshold $p_c$ at low temperature. The critical behavior of this transition is also of conventional power-law type and our Monte Carlo data can be fitted well with theoretical behavior predicted within the scaling theory by Vojta and Schmalian\cite{VojtaSchmalian05b}.  

An experimental realization of the three-dimensional superfluid-Mott glass transition can be found in diluted anisotropic spin-1 antiferromagnets. These systems are typically three-dimensional and exhibit particle-hole symmetry naturally as a consequence of the up-down symmetry of the Hamiltonian in the absence of external magnetic field. Such a realization was recently observed in bromine-doped dichloro-tetakic-thiourea-nickel\cite{Yuetal12}.  

Further experimental studies can be carried out in disordered bosonic systems such as ultracold atoms in optical lattices as well as granular superconductors. However, often only \textit{statistical} particle-hole symmetry can be achieved in these systems. Whether or not this statistical particle-hole symmetry will destabilize the Mott glass into a Bose glass remains still unresolved. 

\begin{acknowledgements}
This work was supported in part by the NSF under Grants No. PHY-$1125915$ and DMR-$1506152$. T.V. acknowledges the hospitality of the Kavli Institute for Theoretical Physics, where part of the work was performed. 
\end{acknowledgements}

\bibliographystyle{apsrev4-1}
\bibliography{rareregions}

\end{document}